\documentclass[11pt]{article}
\usepackage{amsmath}
\usepackage{amsthm}
\usepackage{amssymb}
\usepackage{latexsym}

\newtheorem{thm}{Theorem}[section]
\newtheorem{lemma}{Lemma}[section]
\newtheorem{defn}{Definition}[section]
\newtheorem{remark}{Remark}[section]
\newtheorem{example}{Example}[section]
\newtheorem{cor}{Corollary}[section]

\DeclareMathOperator{\id}{id}
\newcommand{\relby}[1]{\stackrel{#1}{\sim}}

\begin{document}
\title{Classifying all Mutually Unbiased Bases in $\mathbf{Rel}$}
\author{Julia Evans\footnote{School of Computer Science, McGill
    University} \and Ross Duncan\footnote{Computing Laboratory, Oxford University} \and Alex Lang\footnotemark[1] \and Prakash Panangaden\footnotemark[1]}
\date{\today}
\maketitle
\begin{abstract}
  Finding all the mutually unbiased bases in various dimensions is a
  problem of fundamental interest in quantum information theory and pure
  mathematics.  The general problem formulated in finite-dimensional
  Hilbert spaces is open.  In the categorical approach to quantum mechanics
  one can find examples of categories which behave ``like'' the category of
  finite-dimensional Hilbert spaces in various ways but are subtly
  different.  One such category is the category of sets and relations,
  $\mathbf{Rel}$.  One can formulate the concept of mutually unbiased bases
  here as well.  In this note we classify all the mutually unbiased bases
  in this category by relating it to a standard question in combinatorics.
\end{abstract}

\section{Introduction}
In the early 1960s Julian
Schwinger~\cite{Schwinger59,Schwinger60,Schwinger60a,Schwinger60b}
initiated a new approach to the foundations of quantum mechanics by basing
the subject on the algebra of measurements.  A mature presentation of this
approach appears in the recent book~\cite{Schwinger03} published
posthumously.  Schwinger identified a fundamental concept: mutually
unbiased bases which lay at the heart of the algebra of measurements and
the geometry of quantum states.  

A basis in the state space of a quantum system defines a
measurement~\cite{Peres95}.  Two bases are said to be \emph{mutually
  unbiased} if each vector of the first basis has the same inner product
with every vector of the second.  Thus two mutually unbiased bases (MUBs)
define \emph{complementary} observables.  Finding maximum sets of MUBs in a
vector space is a challenging problem in geometry and remains open.  It is
known that there are at most $d+1$ MUBs in spaces of dimension $d$.  It is
possible to achieve this upper bound when $d$ is a prime power; see for
example~\cite{Wocjan04}.  However, for other dimensions little is known.
In particular for dimension $6$ it is known how to construct $3$ MUBs and
numerical evidence suggests that there are no more, but no proof is known.

The recent categorical approach to the foundations of quantum mechanics
initiated by Abramsky and Coecke~\cite{Abramsky04} and pursued by
Selinger~\cite{Selinger06b,Selinger07} and
others~\cite{Coecke08,Coecke08b,Coecke09} gives a number of ``alternative
universes'' in which theories like -- but not identical to -- orthodox
Hilbert-space-based quantum mechanics can be explored.  In particular the
category of sets and binary relations $\mathbf{Rel}$ has many of the
features required for quantum mechanics, but is also clearly different.  In
this note we classify all the possible MUBs in this category.  In
particular, we show that there are only $3$ when the underlying set has
$6^2 = 36$ elements.  This is suggestive of the the situation with
finite-dimensional Hilbert spaces where only 3 mutually unbiased bases are
known for $6$ dimensions.  However, the analogy is not perfect.  The
category $\mathbf{Rel}$ behaves like vector spaces with the scalars being
$\{0,1\}$.  Given a set $S$ of $n$ elements we can regard the elements of
$S$ as basis ``vectors'' and subsets of $S$ can be regarded as formal
linear combinations with coefficients $0$ or $1$.  The set with $36$
elements can be thought of as analogous to the space of $6\times 6$
matrices, which is not exactly the same as the case of $6$ dimensions.

The proofs are based on combinatorial structures called Latin
squares.  The proof hints at connections with group representation theory
that may help resolve the open questions in the category of
finite-dimensional Hilbert spaces.

In the program of categorical quantum mechanics there have been some recent
papers that set the stage for the present work.  First, Coecke, Paquette,
Pavlovic and Vicary~\cite{Coecke08,Coecke08b} have developed a theory of
measurements in the abstract categorical framework.  They have defined a
\emph{classical structure} as a space equipped with a notion of copying and
deleting satisfying some basic algebraic laws.  They have showed that in the
category $\mathbf{FDHilb}$ of finite dimensional complex Hilbert spaces and
linear maps this amounts to choosing a basis in the space.
Later Coecke and Duncan~\cite{Coecke09b} developed a theory of interacting
observables and described a diagrammatic presentation of the algebra of such
pairs of observables.  Such pairs of observables correspond exactly to MUBs.

In a recent paper Pavlovic~\cite{Pavlovic09} classified all the classical
structures in the category $\mathbf{Rel}$.  He showed that all such
structures come from direct sums of finite abelian groups; a more precise
statement appears below.  Roughly speaking, one has to partition a set and
then provide an abelian group structure on each block of the partition.  In
our classification of MUBs it turns out that the partition is crucial but
the abelian group structure chosen is not.    

\section{Background}
We review some of the basic combinatorial background.  We refer to the
papers of Coecke et al. for the background on categorical quantum mechanics,
but we will review the definition of a classical structure.

\begin{defn}
  A $d \times d$ \emph{Latin square} is a $d\times d$ array filled with $d$
  symbols such that each row and column contains exactly $1$ copy of each
  symbol.
\end{defn}

\begin{defn}
    A \emph{partition} of a set $X$ is a set of disjoint sets $\pi$ such that
    $\displaystyle\bigcup_{S \in \pi} S = X$.
\end{defn}

\begin{defn}
    In a dagger symmetric monoidal category, $f: A \to B$ is \emph{unitary} if
    $f \circ f^\dagger =  \id_B$, $f^\dagger \circ f = \id_A$.
\end{defn}
It is easy to see that in $\mathbf{Rel}$, a relation is unitary if and only
if it is a bijective function.

\begin{defn}
  A classical structure in a dagger symmetric monoidal category
  $(\mathbf{C},\otimes,I)$ is a triple $(X,\delta, \varepsilon)$ where $X\in
  \mathbf{C}_0$ is an object in $\mathbf{C}$, $\delta: X\to X\otimes X$ is a
  morphism called the copying operation and $\varepsilon: X\to I$ is called the
  deletion and for which $(X,\delta^\dagger, \varepsilon^\dagger, \delta,
  \varepsilon)$ forms a special Frobenius algebra.
\end{defn}

\begin{defn}
    A point $p: I \to X$ is called \emph{classical} for the classical
    structure $(X,\delta,\epsilon)$ if $\delta \circ p= p\otimes p$.
\end{defn}

\begin{defn}
    For any point $p: I \to X$ and classical structure $(X,\delta,\epsilon)$,
    define $\Lambda(p) = \delta^\dagger \circ (p \otimes \id_X)
    \circ \lambda_X^{-1}: X \to X$. $p$ is called \emph{unbiased} for the
    classical structure $(X,\delta,\epsilon)$ if $\Lambda(p)$ is unitary.
\end{defn}

In $\mathbf{Rel}$, what this means is that a set $U$ is unbiased for the
classical structure $(X,\delta,\epsilon)$ if and only if the relation $R$
defined by 

\[x \relby{R} y \iff \exists \ z \in U \text{ such that } x \relby{\delta} (y,z)\]

is a bijection.

\section{Classical Structures in $\mathbf{Rel}$}

In \cite{Pavlovic09}, Pavlovic showed that 
\begin{thm}
    Every classical structure $(X,\delta,\epsilon)$ in $\mathbf{FRel}$ on a set
    $X$ comes from choosing
\begin{itemize}
    \item a partition of the set $X$
    \item an abelian group operation $\cdot_S$ on every set $S$ in the partition
\end{itemize}
where $\delta$ is defined as follows:\\
$x \relby{\delta} (y,z)$ if $y, z$ are in the same set $S$ in the partition
and $x = y \cdot_S z$.  Here $\epsilon$ is the set of all the group
identities.

Every partition and set of group operations uniquely determine a classical
structure. 
\end{thm}

The following theorem tells us what the classical and unbiased points of a
classical structure are.
\begin{thm}
    \label{unbiased}
    Suppose $(X,\delta, \epsilon)$ is a classical structure with partition
    $\pi$. Then the classical points of $X$ are exactly the sets of $\pi$,
    and the unbiased points are the sets obtained by taking one element
    from each set of $\pi$. 
\end{thm}

Note that the classical and unbiased points do \emph{not} depend on the
group structures chosen on the sets in the partition, only on the partition.

\begin{proof}
    First we will prove that the classical points are the sets in $\pi$.

    Suppose we have a classical point $P$.  Choose $S\in\pi$ such that $|P\cap
    S| \not= \emptyset$.  Let $x\in P \cap S$. 
    For $y,z\in S$, $x\relby{\delta} (y,z)$ if and only if $x = y\cdot_S z$.

    So it follows that $x\relby{\delta}(y,y^{-1}x)$ $\forall$ $y\in
    S$.  Since $\delta \circ P = P \otimes P$, this implies $S\subseteq P$. 

    Now, suppose $\exists$ $y\in P\setminus S$ (that is, $P\neq S$).  Then
    $(x,y)\in P\times P = \delta \circ P$, so $\exists$ $z\in P$ such that
    $z\relby{\delta}(x,y)$.  But this is impossible, since $x$ and $y$ are
    in different sets in the partition.  So $S=P$ and we are done.

    Now we need to show that the unbiased points are the sets obtained by
    taking one element from each set of $\pi$.  Recall that a point $P$ is
    unbiased if and only the relation $R$ defined by 
     \[x \relby{R} y \iff \exists \ z \in P \text{ such that } x
    \relby{\delta} (y,z)\] 
    is a bijection.

    Suppose $S\in \pi$, $x\in S$.  Then the set of things that $x$ is related to
    by $R$ is exactly $\left\{ x \cdot_{S} u^{-1} : u \in S \cap P \right\}$,
    so has cardinality $|S \cap P|$.  So $R$ is a bijection if and only if 
    $|S \cap P| = 1$ for all $S \in \pi$.
\end{proof}
\begin{remark}
    The maps $\Lambda(p)$ with composition form a group isomorphic to the
    direct product of the groups chosen on the sets $S$ in the partition.
\end{remark}
The following corollary will be useful in our analysis of complementary
classical structures. 
\begin{cor}
    \label{samecardinality}
    If $S_1,S_2\subseteq X$ are unbiased points for $(X,\delta, \epsilon)$, then $\left|S_1\right| = \left|S_2\right|$.
\end{cor}
\begin{proof}
    Every unbiased point $S$ is constructed by taking one element from each set
    in the partition, so $\left|S\right|$ is always the number of sets in
    the partition.
\end{proof}

\section{Complementary classical structures}

In this section we give a complete characterisation of complementary classical
structures in $\mathbf{Rel}$ in terms of their partitions, and reduce the
problem of finding complementary classical structures to the well-studied
combinatorial problem of finding mutually orthogonal Latin squares (MOLS).

\begin{defn}
    Two classical structures $(X,\delta,\epsilon)$ and $(X',\delta',\epsilon')$
    are called \emph{complementary} if each classical point of $X$ is an
    unbiased point of $X'$ and vice versa. 
\end{defn}
\begin{defn}
    A partition is \emph{uniform} if all of its parts have the same size.
\end{defn}

\begin{defn}
    Two partitions $\pi_1$ and $\pi_2$ are \emph{complementary} if for every $S \in \pi_1, T\in \pi_2$, $\left|S \cap T\right| = 1$.
\end{defn}

Consider two classical structures on $\left\{ 1,2 \right\}$ with
partitions $\left\{ \left\{ 1 \right\},\left\{ 2 \right\} \right\}, \left\{ \left\{ 1,2 \right\} \right\}$. We will write these partitions as

\begin{center}
    \begin{tabular}{c}
    \hline
    1 \\
    \hline
    2 \\
    \hline
\end{tabular}
,
    \begin{tabular}{cc}
    \hline
    1 & 2\\
    \hline
\end{tabular}. 
\end{center}

respectively. The first has classical points $\left\{ 1 \right\}, \left\{ 2 \right\}$, and
unbiased point $\left\{ 1,2 \right\}$, and the second has classical point
$\left\{ 1,2 \right\}$, and unbiased points $\left\{ 1 \right\}, \left\{ 2
\right\}$. So these two observables are complementary.

Now consider a classical structure $(X,\delta,\epsilon)$ with partition
\[
    \begin{tabular}{cc}
    \hline
    1 & 3\\
    \hline
    2\\
    \cline{1-1}
\end{tabular}
\]

This has classical points $\left\{ 1,3 \right\}, \left\{ 2 \right\}$ and
unbiased points $\left\{ 1,2 \right\}, \left\{ 3,2 \right\}$. We know by
Corollary~\ref{samecardinality} that $\left\{ 1,3 \right\}$ and $\left\{ 2
\right\}$ cannot be unbiased points for the same classical structure since they
have different cardinality, so there is no classical structure complementary to
$X$!  This does not happen in $\mathbf{FDHilb}$, where every observable in  a space of dimension greater than 1 has a
complementary observable.  The same argument shows that any classical structure
that does not have a uniform partition has no complementary classical structure. 

Theorem \ref{unbiased} says that two classical structures are complementary if
and only if their partitions are complementary.

Now, suppose we have a classical structure $X$ with a uniform partition 

\[
\pi = 
\begin{tabular}{cccc}
    \hline
    $\pi_{11}$ & $\pi_{12}$ & \ldots & $\pi_{1k}$ \\
    \hline
    $\pi_{21}$ & $\pi_{22}$ & \ldots & $\pi_{2k}$ \\
    \hline
    \ldots & \ldots & \ldots & \ldots \\
    \hline
    $\pi_{\ell 1}$ & $\pi_{\ell 2}$ & \ldots & $\pi_{\ell k}$ \\
    \hline
\end{tabular}
\]

Then the transpose partition
\[
\tau = 
\begin{tabular}{cccc}
    \hline
    $\pi_{11}$ & $\pi_{21}$ & \ldots & $\pi_{\ell1}$ \\
    \hline
    $\pi_{12}$ & $\pi_{22}$ & \ldots & $\pi_{\ell 2}$ \\
    \hline
    \ldots & \ldots & \ldots & \ldots \\
    \hline
    $\pi_{1k}$ & $\pi_{2k}$ & \ldots & $\pi_{\ell k}$ \\
    \hline
\end{tabular}
\]

is complementary to $\pi$, so if we take a classical structure $X'$ that has
this partition, then $X'$ is complementary to $X$.

So from the previous discussion, we have the following theorem:

\begin{thm}
    A classical structure has a complementary classical structure if and only
    its partition is uniform. Further, two classical structures with
    partitions $T_1$ and $T_2$ are complementary if and only if the partitions
    $T_1$ and $T_2$ are complementary.
\end{thm}

We also have the following corollary:
\begin{cor}
    If a classical structure's partition is not square, then it cannot be part
    of a set of 3 or more mutually complementary classical structures (MCCS). 
\end{cor}

So to have more than $2$ mutually complementary classical structures on $n$
elements, we need $n = d^2$ for some $d$ and also for the classical structures'
partitions to be square. 

Now, let us consider square $d \times d$ partitions. 

\begin{example}
If $d=2$, then we can construct $3$ mutually complementary classical structures on $4$
elements from the following $3$ partitions:

\begin{center}
\begin{tabular}{ccc}

    \begin{tabular}{ccc}
        \hline
        1 & 2 \\
        \hline
        3 & 4  \\
        \hline
    \end{tabular} 
&
    \begin{tabular}{ccc}
        \hline
        1 & 3 \\
        \hline
        2& 4  \\
        \hline
    \end{tabular} 
&
    \begin{tabular}{ccc}
        \hline
        1 & 4 \\
        \hline
        2 & 3  \\
        \hline
    \end{tabular} 
\end{tabular} 
\end{center}
\end{example}

\begin{example}
    \label{3MCCS}
Similarly, if $d=3$, we can construct $4$ mutually complementary classical structures on
$9$ elements from the following $4$ partitions:

\begin{center}
\begin{tabular}{cccc}

    \begin{tabular}{ccc}
        \hline
        1 & 2 & 3 \\
        \hline
        4 & 5 & 6 \\
        \hline
        7 & 8 & 9 \\
        \hline
    \end{tabular} 
&
    \begin{tabular}{ccc}
        \hline
        1 & 4 & 7 \\
        \hline
        2 & 5 & 8 \\
        \hline
        3 & 6 & 9 \\
        \hline
    \end{tabular} 
&
    \begin{tabular}{ccc}
        \hline
        1 & 6 & 8 \\
        \hline
        2 & 4 & 9 \\ 
        \hline
        3 & 5 & 7 \\
        \hline
    \end{tabular} 
&
    \begin{tabular}{ccc}
        \hline
        1 & 5 & 9 \\
        \hline
        2 & 6 & 7 \\
        \hline
        3 & 4 & 8 \\
        \hline
    \end{tabular} 
\end{tabular} 
\end{center}
\end{example}

\begin{lemma}
    From any set of $k$ mutually complementary partitions on a set of $d^2$
    elements we can construct $k-2$ mutually orthogonal Latin squares.
\end{lemma}

\begin{proof}

Suppose we have partitions $\pi_1,\pi_2,\ldots,\pi_k$. Then we can define a table 

\[T = 
\begin{tabular}{|c|c|c|c|}
    \hline
    $T_{11}$ & $T_{12}$ &\ldots & $T_{1d}$ \\
    \hline
    $T_{21}$ & $T_{22} $ & \ldots & $T_{2d}$ \\
    \hline
    \ldots & \ldots & \ldots & \ldots \\
    \hline
    $T_{d1}$ & $T_{d2} $ & \ldots & $T_{dd}$ \\
    \hline
\end{tabular}
\]

such that the rows of $T$ are the parts of $\pi_1$ and the columns of $T$ are
the parts of $\pi_2$. 

Now, for any partition $\pi$ of $X$ complementary to both $\pi_1$ and $\pi_2$, we
can define a Latin square $L_\pi$ by assigning a symbol $\alpha_S$ to each part
$S$ of $\pi$, and letting 

\[L_{\pi_{ij}} = \alpha_S \text{ if and only if } T_{ij} \in S\]

For any such partitions $\sigma$ and $\tau$, $L_\sigma$ and $L_\tau$ are
orthogonal Latin squares if and only if $\sigma$ and $\tau$ are complementary
partitions. 

So from $\pi_1,\pi_2,\ldots,\pi_k$ we get the $k-2$ orthogonal Latin squares
$L_{\pi_3},L_{\pi_4},\ldots,L_{\pi_k}$.
\end{proof}

\begin{example}
    We can obtain the partitions in Example \ref{3MCCS} from the table 
    \begin{center}
         \begin{tabular}{|c|c|c|}
            \hline
            1 & 2 & 3 \\
            \hline
            4 & 5 & 6 \\
            \hline
            7 & 8 & 9 \\
            \hline
        \end{tabular}
    \end{center}
    
    and the orthogonal Latin squares

    \begin{center}
        \begin{tabular}{|c|c|c|}
            \hline
            1 & 2 & 3 \\
            \hline
            2 & 3 & 1 \\
            \hline
            3 & 1 & 2 \\
            \hline
        \end{tabular}
        ,
        \begin{tabular}{|c|c|c|}
            \hline
            1 & 3 & 2 \\
            \hline
            2 & 1 & 3 \\
            \hline
            3 & 2 & 1 \\
            \hline
        \end{tabular}
    \end{center}
\end{example}
We can also prove the converse of this:
\begin{lemma}
    From any set of $k$ mutually orthogonal $d\times d$ Latin squares, we can
    construct $k+2$ mutually complementary partitions of a set $X$ with $d^2$
    elements.
\end{lemma}

\begin{proof}
    Suppose we have $k$ $d \times d$ Latin squares $L^1,\ldots,L^k$. Put the
    elements of $X$ into a $d\times d$ array $T$. 

    Let $\sigma$ and $\tau$ be the partitions of $X$ obtained from the rows and
    columns of $X$ respectively.  

    For each Latin square $L^i$, define the partition 
    
    \[\pi_i = \left\{ \left\{
    T_{jk} : L^i_{jk} = \alpha \right\} : \alpha \text{ a symbol of } L^i
    \right\}\]
    
    Then the partitions $\sigma,\tau,\pi_1,\ldots,\pi_k$ form a set
    of $k+2$ mutually complementary partitions.
\end{proof}

Since two classical structures $X,X'$ are complementary if and only if their
partitions are complementary, we have 

\begin{thm}
    There are $k$ classical structures on a set with $d^2$ elements if and only
    if there exist $k-2$ $d\times d$ mutually orthogonal Latin squares.
\end{thm}

The equivalence of sets of mutually complementary partitions and MOLS can
be found in \cite{Colbourn06}. 

Wocjan and Beth showed that if there are $k$ $d\times d$ Latin squares then
there are $k+2$ MUBS in $\mathbb{C}^{d^2}$ \cite{Wocjan04}, so we have the
following corollary relating $MCCS$ in $\mathbf{FRel}$ and $\mathbf{FDHilb}$.
\begin{cor}
    If there are $k$ MCCS on $d$ elements in $\mathbf{FRel}$, then there
    are $k$ MUBS in $\mathbb{C}^{d^2}$. 
\end{cor}

In particular, we know that there are $d-1$ orthogonal $d \times d$ Latin
squares if $d$ is a prime power, and that there are no pairs of orthogonal
Latin squares on 6 elements \cite{Colbourn06}, so we have the following
corollary (the number of MUBS on 6 elements in $\mathbf{FDHilb}$ is
unknown):

\begin{cor}
    There are at most $d+1$ MCCS on $d^2$ elements. If $d$ is a prime power,
    then there are $d+1$ MCCS on $d^2$ elements. For $d=6$, there are exactly
    $3$ MCCS on $d^2$ elements.
\end{cor}

\section{Conclusions}
The category $\mathbf{Rel}$ is a ``toy'' version of quantum mechanics where
one can explore ideas in a simpler setting.  In our proofs the structures
that appear are partitions of sets and Latin squares.  It suggests that for
Hilbert spaces the relevant structures might be Young tableaux though, at
present, we have no idea how to pursue this thought.  Another intruiging
link with algebra comes from that fact that Latin squares are the
multiplication tables of loops (non-associative analogues of groups).

The proofs depend on specific properties of $\mathbf{Rel}$; this suggests
that the abstract diagrammatic algebra, while useful for general results,
will not help with tackling the problem of classifying MUBs in
$\mathbf{FDHilb}$. 

\section*{Acknowledgements} This research was suppored by a grant from the
Office of Naval Research.
\bibliography{../../../Documents/Bib/main}

\begin{thebibliography}{BCV08}

\bibitem[AC04]{Abramsky04}
S.~Abramsky and B.~Coecke.
\newblock A categorical semantics of quantum protocols.
\newblock In {\em Proceedings of the 19th Annual IEEE Symposium on Logic in
  Computer Science: LICS 2004}, pages 415--425. IEEE Computer Society, 2004.

\bibitem[BCV08]{Coecke08}
Dusko~Pavlovic Bob~Coecke and Jamie Vicary.
\newblock A new description of orthogonal bases.
\newblock Available from \verb+http://arxiv.org/abs/0810.0812+, 2008.

\bibitem[CD06]{Colbourn06}
Charles~J. Colbourn and Jeffrey~H. Dinitz, editors.
\newblock {\em Handbook of Combinatorial Designs}.
\newblock Chapman \& Hall, 2006.

\bibitem[CD09]{Coecke09b}
Bob Coecke and Ross Duncan.
\newblock Interacting quantum observables: categorical algebra and
  diagrammatics.
\newblock \verb+arXiv quant-ph 0906.4725+, 2009.

\bibitem[CES09]{Coecke09}
Bob Coecke, Bill Edwards, and Rob Spekkens.
\newblock The group theoretic origin of non-locality for qubits.
\newblock Technical Report RR-09-04, OUCL, 2009.

\bibitem[CPP08]{Coecke08b}
Bob Coecke, Eric~O Paquette, and Dusko Pavlovic.
\newblock Classical and quantum structures.
\newblock Technical Report RR-08-02, OUCL, 2008.

\bibitem[Pav09]{Pavlovic09}
Dusko Pavlovic.
\newblock Quantum and classical structures in nondeterministic computation.
\newblock In Peter Bruza, Don Sofge, and Keith {van Rijsbergen}, editors, {\em
  Proceedings of Quanum Interaction 2009}, Lecture Notes in Computer Science.
  Springer Verlag, 2009.
\newblock 15 pp.

\bibitem[Per95]{Peres95}
A.~Peres.
\newblock {\em Quantum Theory: Concepts and Methods}.
\newblock Kluwer Academic, 1995.

\bibitem[Sch59]{Schwinger59}
Julian Schwinger.
\newblock The algebra of microscopic measurement.
\newblock {\em Proc. Natl. Acad. Sci.}, 45(10):1542--1553, 1959.

\bibitem[Sch60a]{Schwinger60}
Julian Schwinger.
\newblock The geometry of quantum states.
\newblock {\em Proc. Natl. Acad. Sci.}, 46:257--265, 1960.

\bibitem[Sch60b]{Schwinger60b}
Julian Schwinger.
\newblock The special canonical group.
\newblock {\em Proc. Nat. Acad. Sci.}, 46:1401--1415, 1960.

\bibitem[Sch60c]{Schwinger60a}
Julian Schwinger.
\newblock Unitary operator bases.
\newblock {\em Proc. Natl. Acad. Sci.}, 46:570--579, 1960.

\bibitem[Sch03]{Schwinger03}
Julian Schwinger.
\newblock {\em Quantum mechanics: symbolism of atomic measurements}.
\newblock Springer-Verlag, 2003.
\newblock Edited by Bertolt-Georg Engelert.

\bibitem[Sel07]{Selinger07}
Peter Selinger.
\newblock Dagger compact closed categories and completely positive maps.
\newblock In {\em In Proceedings of the 3rd International Workshop on Quantum
  Programming Languages (QPL 2005)}, number 170 in ENTCS, pages 139--163, 2007.

\bibitem[Sel08]{Selinger06b}
Peter Selinger.
\newblock Idempotents in dagger categories.
\newblock In {\em Proceedings of the 4th International Workshop on Quantum
  Programming Languages Oxford 2006}, volume 210 of {\em ENTCS}, pages
  107--122, 2008.

\bibitem[WB04]{Wocjan04}
Pawel Wocjan and Thomas Beth.
\newblock New construction of mutually unbiased bases in square dimensions.
\newblock Available from
  \verb+http://www.citebase.org/abstract?id=oai:arXiv.org:quant-ph/0407081+,
  2004.

\end{thebibliography}
\end{document}